\documentclass[reprint,amsmath,amssymb,aps,superscriptaddress]{revtex4-2}
\usepackage{dcolumn}
\usepackage{mathrsfs}
\usepackage{amsmath,gensymb}
\usepackage{amsfonts}
\usepackage{amssymb}
\usepackage{amsthm}
\usepackage{graphicx}
\usepackage{natbib}
\usepackage{color}
\usepackage{hyperref}
\usepackage{bm}
\usepackage[caption=false]{subfig}
\usepackage{verbatim}
\RequirePackage{lineno}
\providecommand{\U}[1]{\protect\rule{.1in}{.1in}}

\begin{document}
\title{Selective Amplification of the Topological Hall Signal in Cr$_2$Te$_3$: The Role of Molecular Exchange Coupling}
\author{Suman Mundlia}
\affiliation{Tata Institute of Fundamental Research, Hyderabad, Telangana 500046, India}
\author{Ritesh Kumar}
\affiliation{Tata Institute of Fundamental Research, Hyderabad, Telangana 500046, India}
\author{Anshika Mishra}
\affiliation{Tata Institute of Fundamental Research, Hyderabad, Telangana 500046, India}
\author{Malavika Chandrasekhar}
\affiliation{Tata Institute of Fundamental Research, Hyderabad, Telangana 500046, India}
\author{Narayan Mohanta}
\affiliation{Department of Physics, Indian Institute of Technology Roorkee, Roorkee 247667, India}
\author{Karthik V. Raman}
\email{kvraman@tifrh.res.in}
\affiliation{Tata Institute of Fundamental Research, Hyderabad, Telangana 500046, India}

\begin{abstract} 
Layered magnetic transition-metal chalcogenides (TMCs) are a focal point of research, revealing a variety of intriguing magnetic and topological ground states. Within this family of TMCs, chromium telluride has garnered significant attention because of its excellent tunability in magnetic response, owing to the presence of competing magnetic exchange interactions. We here demonstrate the manipulation of magnetic anisotropy in ultra-thin Cr$_2$Te$_3$ films through growth engineering leading to a controlled transition from in-plane to out-of-plane orientation with an intermediate non-coplanar magnetic ground phase characterized by a topological Hall effect. Moreover, interfacing these films with Vanadyl phthalocyanine (VOPc) molecules prominently enhances the non-coplanar magnetic phase, attributing its presence to the competing interfacial magnetic exchange interactions over the spin-orbit-driven interfacial effects. These findings pave the way for the realization of novel topological spintronic devices through interface-modulated exchange coupling.

\end{abstract}
\maketitle
\section{Introduction}
\par Layered transition-metal chalcogenides (TMCs) exhibit a rich variety of physical phenomena, including superconductivity \cite{saito2016highly, yan2019nbsete}, charge density waves \cite{hor2005nanowires, hwang2024charge}, ferroelectricity \cite{wang2023towards}, and ferromagnetism \cite{gong2017discovery, deng2018gate}, arising from the intricate interplay of magnetic configuration, real-space, and momentum-space topology. In their ferromagnetic phase, TMCs demonstrate a robust magnetic ground state even down to a monolayer \cite{burch2018magnetism} and host unconventional spin textures such as magnetic skyrmions \cite{nagaosa2013topological,fert2017magnetic, luo2020skyrmion}. In particular, the chromium telluride (CT) family of compounds, such as CrTe$_2$, tr-Cr$_5$Te$_8$, Cr$_2$Te$_3$, Cr$_3$Te$_4$, and CrTe, can potentially exhibit a variety of these magnetic phases, making them promising candidates for spintronic applications \cite{YANG2023106567}.

\par
\begin{figure*}[t]
\centering
\includegraphics[width=17cm]{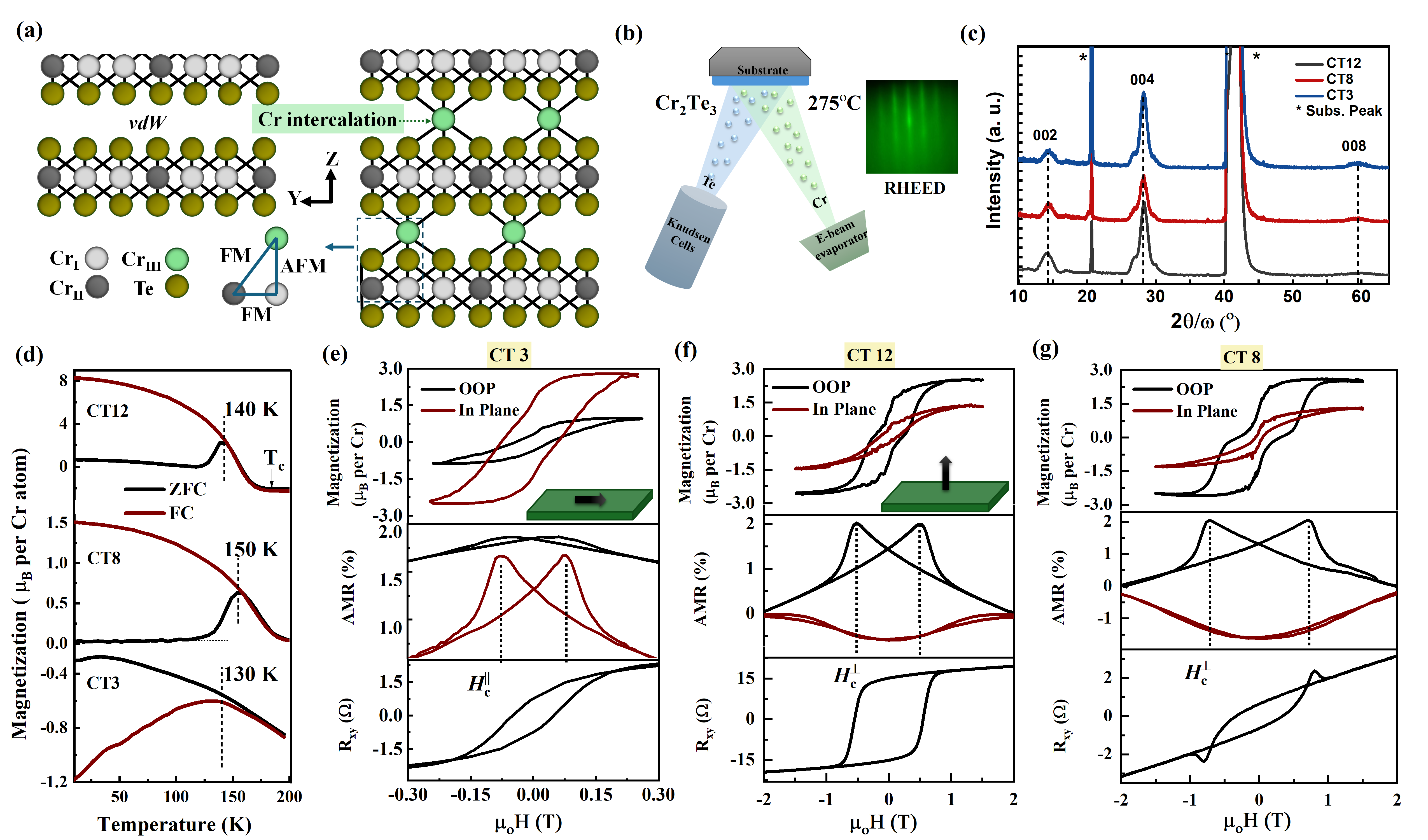}
\caption{(a) Crystal structure of CrTe$_2$ (left) and Cr$_2$Te$_3$ (right) in YZ plane, showcasing three distinct types of Cr atoms (Cr$_{I,II,III}$) with the co-existence of FM and AFM exchange interactions in Cr$_2$Te$_3$. (b) Schematic of the molecular beam epitaxy co-evaporation setup for Cr$_2Te_3$ thin films, with in-situ RHEED monitoring. RHEED image of the 4 nm Cr$_2$Te$_3$ film taken along [10$\bar{1}$0] direction. (c) Coupled XRD scan ($2\theta/\omega$) of CT3, CT8 and CT12 films showing c-axis oriented growth. (d) Magnetization (M) vs Temperature in zero-field cooled (black) and field-cooled (red) procedure for CT3, CT8 and CT12 films. $M$ vs $H$ (top), AMR (in \%) vs $H$ (middle) and Hall resistance (R$_{xy}$) vs $H$ (bottom) for (e) CT3, (f) CT12 and (g) CT8 devices taken at 6 K. Typical zero-field resistance values of the devices were in the range of 2 to 6 k$\Omega$ with the CT3 devices being more resistive.}
\label{fig1}
\end{figure*}

 
Of these compounds, CrTe$_2$ and Cr$_2$Te$_3$ attract significant research interest. CrTe$_2$ crystallizes in a trigonal structure characterized by hexagonal Cr layers intercalated between Te bilayers (see Fig.~\ref{fig1}(a)) and held together by van der Waals interactions \cite{acs.jpclett.1c01901, zhang2021room}. On the other hand, Cr$_2$Te$_3$ is formed by the self-intercalation of a half-occupied Cr layer within the Te bilayers of CrTe$_2$. This results in a metal-deficient NiAs-type crystal structure of Cr$_2$Te$_3$ with hexagonal packing of Te atoms and octahedral sites occupied by Cr atoms with three types of magnetic centres: Cr$_{I}$, Cr$_{II}$, and Cr$_{III}$ \cite{bian2021covalent}. Density functional theory calculations reveal magnetic moments of 3.045, 3.126 and 3.096 Bohr magneton ($\mu_B$) for Cr$_{I}$, Cr$_{II}$, and Cr$_{III}$, respectively \cite{bian2021covalent}.
Similar to CrTe$_2$, the Cr$_{I}$ and Cr$_{II}$ atoms form a ferromagnetic sublattice in the XY plane, while the Cr$_{III}$ atoms sparsely distribute within the self-intercalated layer, exhibiting competing antiferromagnetic and ferromagnetic exchange interactions with Cr$_{I}$ and Cr$_{II}$ atoms, respectively. These competing exchange interactions, sensitive to the c-axis spacing \cite{Li2019Molecular}, result in spin frustration and canting, which are highly sensitive to the Cr intercalation level \cite{hamasaki1975neutron,JDijkstra_1989}. This self-intercalant nature of the Cr layer enables structural stability across a wide range of Cr concentrations, allowing for the tuning of exotic magnetic ground states.

\section{EXPERIMENTAL SECTION}
\subsection{Film growth \& Characterization}
\par In this work, we stabilize the Cr$_2$Te$_3$ phase in approximately 4 monolayer (ML) thick ultra-thin films and investigate its magnetic response resulting from weak modulations in Cr intercalation levels. These modulations were achieved in our molecular beam epitaxy system by varying the relative flux of Cr and Te atomic beams at the optimized growth conditions of the Cr$_2$Te$_3$ phase (see Fig.~\ref{fig1}(b)). We here present results for Cr:Te flux ratios of 1:3, 1:8, and 1:12, labeled CT3, CT8, and CT12, respectively. These films were deposited on c-cut sapphire substrates at 275$^{o}$C, exhibiting predominantly 2D growth revealed by in-situ reflection high-energy electron diffraction (RHEED) studies. Following growth, the films were annealed at 275 $^{o}$C for 30 minutes and then cooled down to room temperature before adding a 3 nm Te capping layer. High-resolution X-ray diffraction (see Fig.~\ref{fig1}(c)) consistently confirmed the Cr$_2Te_3$ phase across varying flux ratios. All samples exhibited similar lattice constants: 11.96 Å for the c-axis and 6.2 Å in-plane, both slightly below the bulk values of 12.07 Å and 6.814 Å, respectively \cite{bian2021covalent, hamasaki1975neutron}. Energy dispersive X-ray analysis (on samples with Al$_2$O$_3$ capping) indicated no observable variations in stoichiometry from the expected 2:3 ratio. Temperature-dependent magnetization measurements (see Fig.~\ref{fig1}(d)) show Curie temperatures ($T_{c}$) near the bulk value of 180 K \cite{Li2019Molecular}, which are considerably lower than the expected T$_c$ values of CrTe$_2$ ($\sim 305$ K) and CrTe phase ($\sim 330$ K)  \cite{zhang2020tunable, meng2021anomalous,LUO2025105779,fujisawa2020tailoring,JDijkstra_1989}. Field-cooled (FC) and zero-field cooling (ZFC) magnetization scans reveal the presence of a blocking temperature ($T_B$) that is associated with the emergence of antiferromagnetic exchange interactions leading to the bifurcation of the FC and ZFC magnetization curves. The variation in $T_B$ is likely caused by subtle variations in the Cr intercalation levels. 
\subsection{Device Characterization}
\par The as-grown films were patterned into Hall bar structures of length $\sim$5 mm and width $\sim$0.4 mm and cooled to 5 K for magnetotransport measurements. Figure ~\ref{fig1}(e)-(g) shows the anisotropic magnetoresistance (AMR) response of devices with different Cr:Te ratios for both in-plane (IP) and out-of-plane (OOP) configurations of magnetic field (H). CT3 devices exhibit distinct AMR peaks at the coercivity field (H$_c$) and a larger saturation magnetization in the IP configuration, indicating a preferred in-plane magnetic anisotropy. In contrast, CT12 devices show a distinct AMR peak, a larger saturation magnetization, and a square-like Hall response in the OOP configuration, thereby exhibiting perpendicular magnetic anisotropy (PMA). This trend is consistent with magnetic anisotropy studies in the extreme cases of CrTe$_2$ phase (low Cr content) \cite{meng2021anomalous} and CrTe phase (high Cr content) \cite{LUO2025105779} showing PMA and IP anisotropy, respectively.

\begin{figure*}[t]
\centering
\includegraphics[width=9cm]{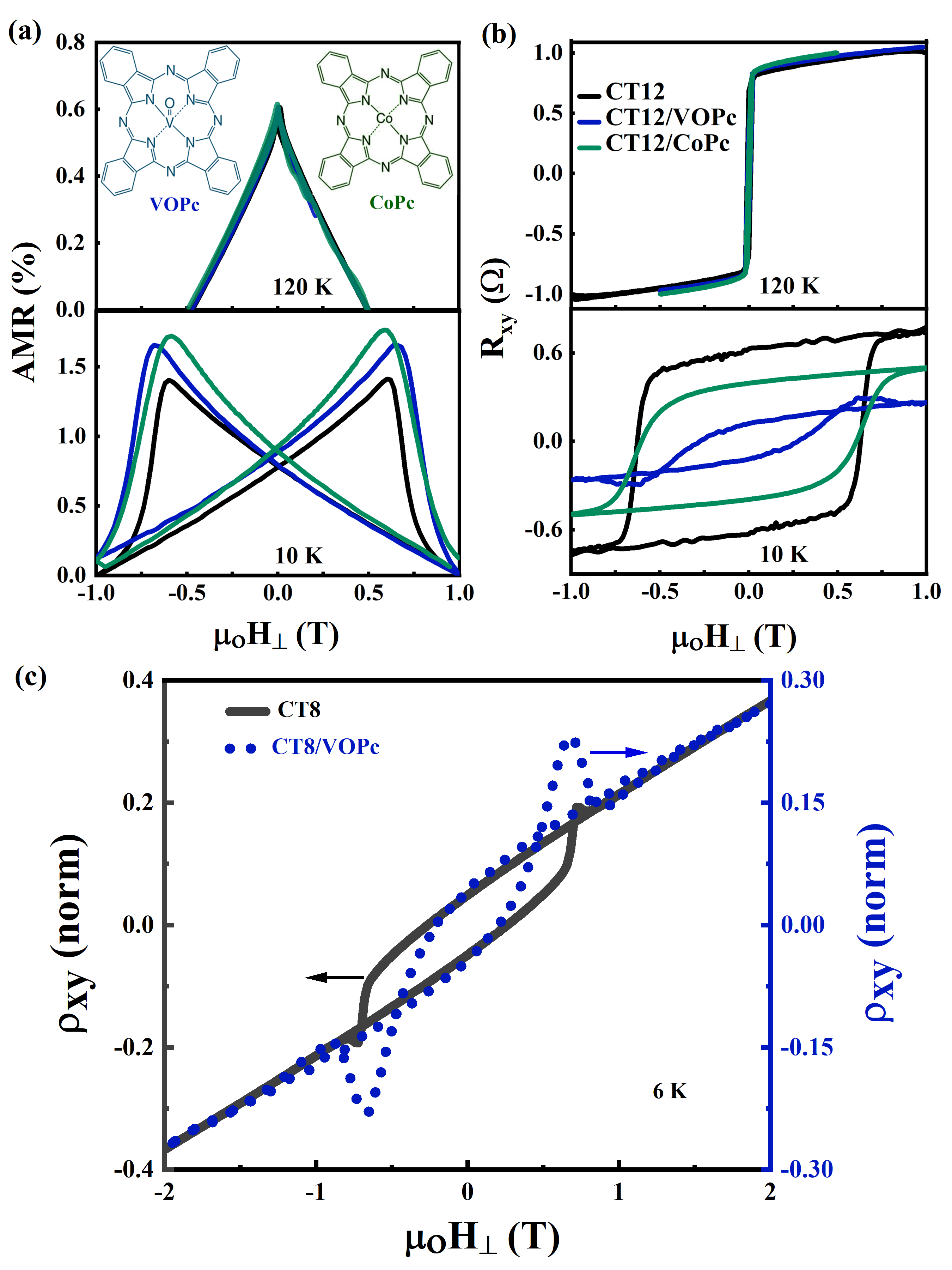}
\caption{(a) AMR (in \%) vs $H$ and (b) Hall signal ($R_{xy}$) vs $H$ of reference CT12 (4.5 nm) (black), CT12 (4.5 nm)/VOPc (4 ML) (blue) and CT12 (4.5 nm)/CoPc (4 ML) (green) devices, at 120 K (top) and 10 K(bottom). (c) Normalized Hall resistivity ($\rho$$_{xy}$(norm)) vs $H$ of reference CT8 (4.5 nm) (black) and CT8 (4.5 nm)/VoPc (4 ML) (blue) devices at 6 K. The Hall resistivity is normalized by $\rho$$_{xy}$(norm) = $\rho$$_{xy}$(H, T)/$\rho$$_{xy}$(1 T, 120 K).}
\label{fig2}
\end{figure*}

\par Figure ~\ref{fig1}(g) shows the response of the CT8 device at the intermediate flux ratio. While the AMR signal suggests PMA, the out-of-plane magnetization response exhibit a sheared hysteresis loop near zero field. Further, these devices show an additional hump-like Hall signal over the conventional anomalous Hall signal ($\text{R}_{xy}$) \cite{RevModPhys.82.1539}, observed over a wider range of temperature spanning 30 K to 2 K. Careful analysis, as documented in the supplementary information, rules out disorder-induced or strain-induced Berry curvature effects as the cause of this response \cite{Chi2023,kimbell2022challenges}. In Cr$_2$Te$_3$, optimally filled self-intercalated Cr layers can induce spin frustration and tilting of magnetic moments \cite{hamasaki1975neutron,JDijkstra_1989}. This configuration does not strongly support IP anisotropy or PMA, similar to what is observed in CT8. Consequently, a complex domain state might arise, potentially hosting non-coplanar magnetic states \cite{zhao2018observation}. The observation of a sheared magnetization response near the zero field in CT8 is also indicative of the presence of compensating exchange interactions with the complex domain state \cite{soumyanarayanan2017tunable,cheng2023room}. This could explain the additional hump-like characteristic that originates from the topological Hall effect (THE)\cite{nagaosa2013topological}. Recent Lorentz TEM studies also indicate the existence of skyrmion states in CrTe$_{1+x}$, supporting our observations in this class of materials \cite{Saha2022}.
\begin{figure*}[htb!]
\centering
\includegraphics[width=11cm]{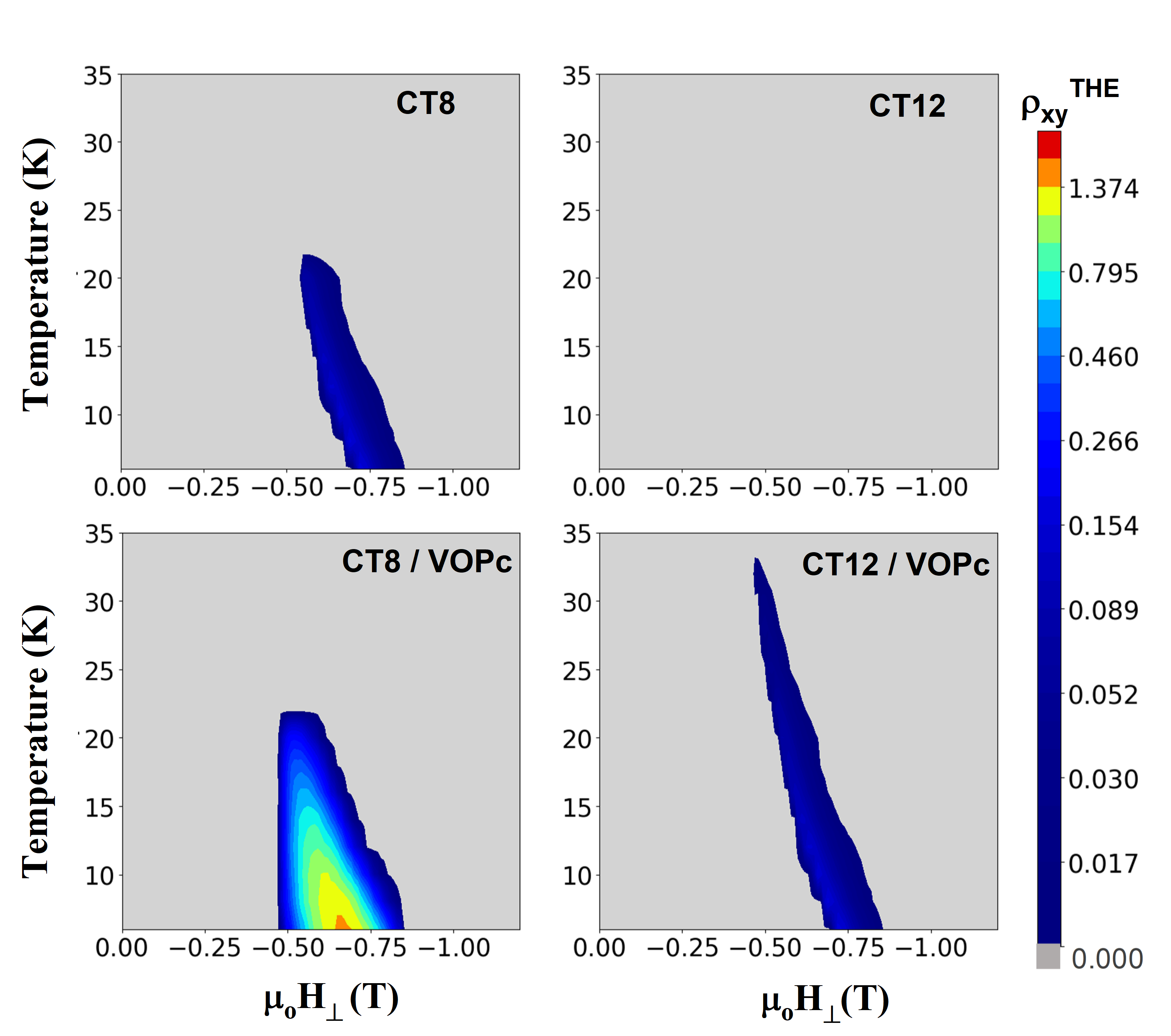}
\caption{3D Color map of $\rho$$_{xy}$(THE) with varying $\text{T}$ and $\text{H}$ for CT8 (4.5 nm), CT12 (4.5 nm), CT8 (4.5 nm)/VOPc (4ML), and CT12(4.5 nm)/VoPc(4 ML) devices. $\rho$$_{xy}$(THE) is obtained by removing the background anomalous Hall signal from $\rho$$_{xy}$. The values of $\rho$$_{xy}$ are in units of 1$0^{-7}$ $\Omega$.cm.}
\label{fig3}
\end{figure*}
\par Contrary to the general understanding that the THE signal in thin films arises from interfacial effects driven by inversion symmetry breaking and high spin-orbit coupling \cite{nagaosa2013topological}, our work decisively reveals an intrinsic origin for the THE in pristine ultra-thin $\text{Cr}_2\text{Te}_3$ films. We demonstrate that this intrinsic effect is controlled by the modulation of competing magnetic exchange interactions via $\text{Cr}$ intercalation. This is supported by the observation of a relatively higher value of T$_B$ in $\text{CT8}$ compared to $\text{CT12}$ or $\text{CT3}$ (see Fig.~\ref{fig1}(d)), confirming the enhanced antiferromagnetic exchange interactions induced by the $\text{Cr}$ intercalation process. We further validate the significance of these interactions through molecular coupling \cite{Raman2014}, specifically by examining the effect of the adsorption of low-spin-orbit metal-phthalocyanine (MPc) molecules, \emph{viz.} cobalt phthalocyanine (CoPc) and vanadyl oxy-phthalocyanine (VOPc), at the $\text{Cr}_2\text{Te}_3$ film surface. Previous studies have demonstrated that the surface absorption of molecules can modify the magnetic anisotropy of a ferromagnetic surface \cite{PhysRevLett.114.247203,PhysRevLett.111.106805,Raman2013}. In this work, we show that such molecular coupling can stabilize a non-coplanar magnetic phase. This stabilization is driven by the selective modulation of interfacial magnetic exchange interactions. To ensure identical growth parameters for the $\text{Cr}_2\text{Te}_3$ layer, a shadow mask technique was utilized. This technique allowed us to simultaneously prepare the reference $\text{Cr}_2\text{Te}_3$ and the $\text{Cr}_2\text{Te}_3/\text{MPc}$ samples in a single growth process.

\par Figure ~\ref{fig2}(a) \& ~\ref{fig2}(b) presents the AMR and Hall responses of the reference CT12, CT12/VOPc(4ML) and CT12/CoPc(4ML) devices at 120 K and 10 K. The Hall signal was normalized to $\rho$$_{xy}$(1 T, 120 K). At 120 K, the AMR and Hall responses of the CT12/CoPc and CT12/VOPc devices are similar to the reference CT12 device, suggesting negligible effects of molecular adsorption. However, differences begin to emerge upon cooling (see Supplementary Information). At 10 K, the AMR signal for the CT12/CoPc and CT12/VOPc devices shows a marginal increase, while the Hall signal exhibits significant changes compared to the reference CT12 devices. Specifically, the CT12/CoPc device displays a marginally weaker Hall signal, while the CT12/VOPc device shows a significantly reduced Hall signal accompanied by a distinct THE signal. In a similar investigation for the CT8 devices, the CT8/VOPc device exhibited a further softened Hall response with an enhanced THE signal (see Fig.~\ref{fig2} (c)), while the CT8 / CoPc device did not show a significant change in THE signal compared to the reference CT8 devices.


\par 
Figure ~\ref{fig3} illustrates the 3D contour map showing the temperature and magnetic field dependence of the THE signal (after subtracting the anomalous Hall signal) for the CT8 and CT12 devices, comparing them with and without VOPc adsorption. In CT12/VOPc devices, the THE signal emerges below 20 K and intensifies as the temperature decreases. In CT8/VOPc devices, a more intense THE signal appears over a broader temperature range compared to the reference CT8 devices. These results indicate a preferential modulation of THE response by VOPc adsorption compared to CoPc adsorption, with the VOPc adsorption stabilizing the complex magnetic structure over a wider range of magnetic-field strength. This preferential modulation underscores the crucial role of the central metal ion in tuning surface magnetic exchange interactions in Cr$_2$Te$_3$ films. Previous studies in these MPc molecules indicate that the vanadyl metal ion-center favors an antiferromagnetic interaction with the magnetic substrate \cite{eguchi2014magnetic}, in contrast to the ferromagnetic interaction of the cobalt metal-ion \cite{annese2011control}. In our study, the Te-terminated $\text{Cr}_2\text{Te}_3$ layer is expected to lead to a weaker interaction with the top MPc molecule. While identifying the exact interactions in Cr$_2$Te$_3$/VOPc and Cr$_2$Te$_3$/CoPc films requires further investigation, these reports strongly imply the existence of distinct interaction mechanisms mediated by the two different metal-ion centers.
\begin{figure*}[t]
\begin{center}
\vspace{-0mm}
\includegraphics[width=160mm]{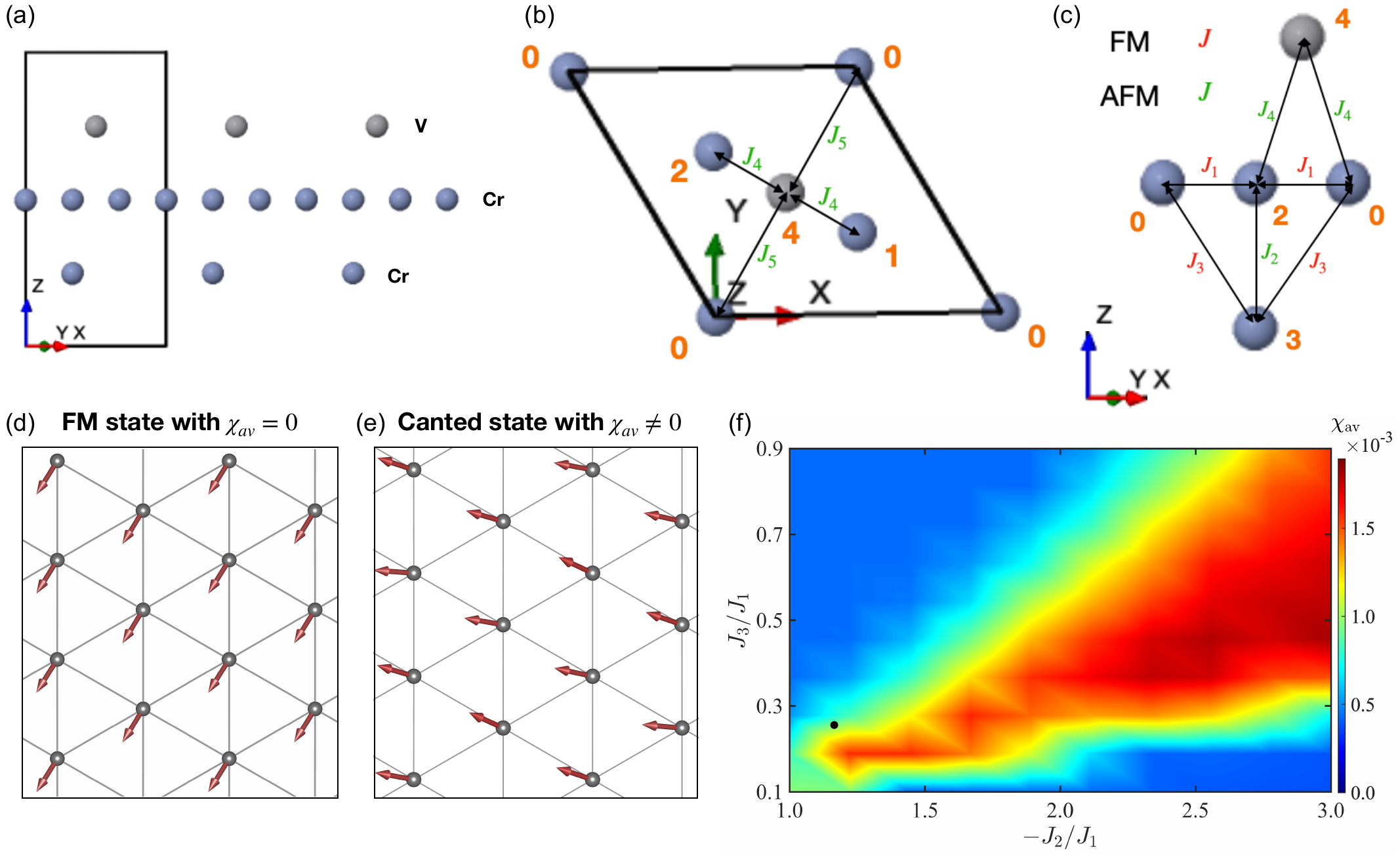}
\caption{(a) Three-layer structure of the magnetic atoms near the considered Cr$_2$Te$_3$/V interface. Moments of Cr and V atoms were taken as 3 $\mu_B$ and 1.72 $\mu_B$, respectively. (b)-(c) Top view and side view of the system, showing different types of exchange interactions considered in the model. (d)-(e) Ground state spin configuration in the top Cr layer, obtained in Monte Carlo annealing calculation, for parameter sets (d) $J_2=0$, $J_3=0$, $J_4=0$, $J_5=0$, and (e) $J_2=-2J_1$, $J_3=0.2J_1$, $J_4=0$, $J_5=0$. In (d), we realize a ferromagnetic (FM) state. In (e), in the presence of finite $J_2$ and $J_3$, a canted state appears, for which the scalar spin chirality is non-zero. (f) Variation of the site-averaged scalar spin chirality $\chi_{\rm av}$ with $J_2$ and $J_3$, revealing that $\chi_{\rm av}$ is enhanced in a parameter regime. The values of $J_4$ and $J_5$ are kept at zero. The black dot in (f) represents the values of $J_2$ and $J_3$, close to the previously-reported parameters for this materials system~\cite{bian2021covalent}. The considered value of $J_1$ is 2.92~meV.}
\label{Fig4}
\vspace{-4mm}
\end{center}
\end{figure*}
\begin{figure*}[t]
\begin{center}
\vspace{-0mm}
\includegraphics[width=90mm]{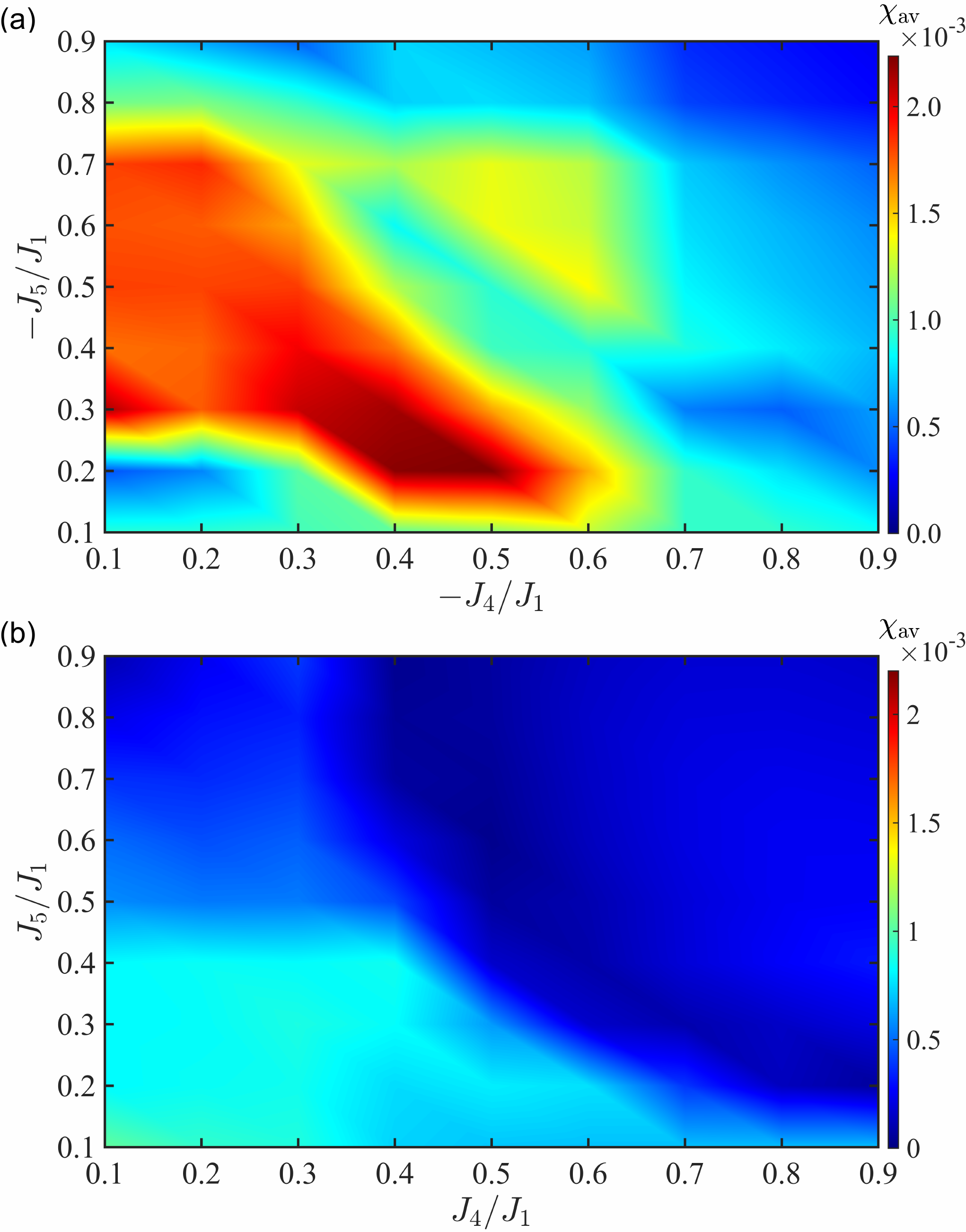}
\caption{(a) Variation of the site-averaged scalar spin chirality $\chi_{\rm av}$ with the exchange couplings $J_4$ and $J_5$ of the V atoms with the interfacial Cr atoms. Both $J_4$ and $J_5$ are considered ferromagnetic, revealing that $\chi_{\rm av}$ is enhanced in a parameter range. The other parameters used are $J_2=-1.2J_1$, $J_3=0.26J_1$, which are close to the previously-reported parameters for this materials system~\cite{bian2021covalent}. (b) Variation of $\chi_{\rm av}$ with $J_4$ and $J_5$, both considered here as antiferromagnetic, also showing that $\chi_{\rm av}$ is enhanced in a range of $J_4$ and $J_5$.
}
\label{Fig5}
\vspace{-4mm}
\end{center}
\end{figure*}

\subsection{Computational Toy Model}
\par To investigate the possibility of THE in the Cr$_2$Te$_3$/VOPc system, originating from scalar spin chirality $\chi={\bf{S}}_i\cdot({\bf{S}}_j \times {\bf{S}}_k)$, where ${\bf{S}}_i$, ${\bf{S}}_j$ and ${\bf{S}}_k$ are spin vectors at three sites $i$, $j$ and $k$ in a triangular plaquette, we develop a theoretical model that incorporates the interfacial exchange interactions due to V atoms on the magnetic state of Cr$_2$Te$_3$. We consider a three-layer system consisting of the magnetic atoms at Cr$_2$Te$_3$/V interface, as shown in Fig.~\ref{Fig4}(a), and use a spin-exchange Hamiltonian ${\cal H}=-\sum_{\langle ij \rangle} J_{ij} {\bf{S}}_i \cdot {\bf{S}}_j$ to describe the magnetic interactions, where $J_{ij}$ represents the strength of the nearest-neighbor exchange interaction between two Cr atoms or between a Cr atom and a V atom. A Cr$_2$Te$_3$ lattice is considered with $J_1$, $J_2$ and $J_3$ as the exchange interaction strengths between the Cr$_{\rm I}$, Cr$_{\rm II}$ and Cr$_{\rm III}$ atoms, as described in Figs.~\ref{Fig4}(b)-(c). For the Cr$_2$Te$_3$ lattice, $J_{1,2} >0$ and $J_3 < 0$ \cite{bian2021covalent}. The terms $J_4$ and $J_5$ describe the exchange interactions between the sparsely populated V layer and the interfacial Cr layer.

\par The ground-state spin configurations at the three layers were obtained using Monte Carlo calculations. We considered a lattice of size 20$\times$20 unit cell (see Fig. 4a), with periodic boundary conditions along the $x$ and $y$ directions. Starting with a completely random spin configuration at a high temperature of 10$J_1$, the temperature is slowly reduced in 1000 steps to a low value of 0.0001$J_1$. At each temperature, 10$^6$ Monte Carlo spin-update steps are performed; in each such step, the total energy of the system, obtained from the above Hamiltonian, is minimized using the Metropolis algorithm. In each spin update step, a new spin direction is chosen randomly within a small cone that spanned the initial spin direction. The average scalar spin chirality is then computed using $\chi_{\rm av}=(1/N)\sum_{\langle ijk \rangle}{\bf{S}}_i\cdot({\bf{S}}_j \times {\bf{S}}_k)$ for the spin configuration at the middle layer of our trilayer system\textit{, i.e.,} the interfacial Cr layer, where $N$ is the total number of lattice sites and $\langle ijk \rangle$ represents three lattice sites in a triangular plaquette. A ferromagnetic ground state is observed in Fig.~\ref{Fig4}(d) for the spin configuration where only the $J_1$ term is non-zero, with all other terms set to zero. When $J_2$ and $J_3$ are switched on, a canted state appears that exhibits a nonzero $\chi_{\rm av}$, as shown in Fig.~\ref{Fig4}(e). The variation of $\chi_{\rm av}$ with $J_2 <0$ and $J_3>0$, depicted in Fig.~\ref{Fig4}(f), indicates that the scalar spin chirality is enhanced by tuning the exchange coupling between Cr$_{\rm I}$, Cr$_{\rm II}$ and Cr$_{\rm III}$ atoms. Experimentally, these exchange interactions can be tuned by altering the level of Cr intercalation, as is done in our samples.

\par Incorporation of V adatoms on top of Cr$_2$Te$_3$ can change the topological Hall response $\chi_{\rm av}$, as shown in Fig.~\ref{Fig5}(a) and Fig.~\ref{Fig5}(b), when the exchange couplings are antiferromagnetic and ferromagnetic, respectively. For the chosen set of parameters $J_2$ and $J_3$, close to the previously-reported values for Cr$_2$Te$_3$ systems~\cite{bian2021covalent} (shown as solid black circle in Fig.~\ref{Fig4}(f)), antiferromagnetic exchange coupling (\textit{i.e.} $J_4$ and $J_5$ negative as in Fig.~\ref{Fig5}(a)) enhances $\chi_{\rm av}$ within a range of parameters; while ferromagnetic exchange coupling (\textit{i.e.} $J_4$ and $J_5$ positive as in Fig.~\ref{Fig5}(b)) tends to reduce $\chi_{\rm av}$. This trend is consistent with our experimental observations of the THE modulation between CT/VoPc and CT/COPc devices, \emph{i.e.} an AFM (FM) interaction between V(Co) and CT surface enhances (weakens) the THE signal.
\section{Conclusion}
\par In this work, we stabilized non-coplanar magnetic textures in ultra thin films of Cr$_2Te_3$. Our method involved carefully controlled chromium intercalation and selective molecular adsorption, driven by specific magnetic exchange interactions at the interface. Our work introduces a paradigm shift by highlighting the critical role of exchange-driven, interface-modulated effects in the formation of novel 2D magnetic textures. Rather than using conventional spin-orbit coupling or interfacial DMI, we show that engineering and modulating competing exchange interactions at the surfaces and interfaces can selectively amplify the topological Hall signal. Together with strain engineering \cite{PhysRevMaterials.7.064002}, this work can open up new avenues to realize future topological spintronics devices. 
\newline
\noindent
\textbf{\\Acknowledgements}: The work was supported by intramural funding at TIFR-Hyderabad from the Department of Atomic Energy, Government of India, under Project Identification No. RTI 4007, external funding from ONRG Grant No. N62909-23-1-2049, DST Nanomission Grant No. DST/NM/TUE/QM-9/2019 (G), and SERB CRG Grant No. CRG/2019/003810. 

%

\end{document}